\documentclass{nature}

\usepackage{graphicx}	
\usepackage{amsmath}	
\usepackage{amssymb}	

\usepackage{hyperref}
\usepackage{url}
\usepackage{microtype}
\usepackage{rotating}
\usepackage{booktabs}
\usepackage{threeparttable}
\usepackage{tabularx}
\title{Hack Weeks as a model for Data Science Education and Collaboration}

\author{Daniela Huppenkothen$^1$, Anthony Arendt$^2$, David W. Hogg$^1$, Karthik Ram$^3$, Jake VanderPlas$^2$, and Ariel Rokem$^2$}

\makeatletter
\let\saved@includegraphics\includegraphics
\AtBeginDocument{\let\includegraphics\saved@includegraphics}
\renewenvironment*{figure}{\@float{figure}}{\end@float}
\makeatother

\begin{document}

\maketitle

\begin{affiliations}
 \item Center for Data Science, New York University
 \item The University of Washington eScience Institute
 \item The Berkeley Institute for Data Science, UC Berkeley
\end{affiliations}

\begin{abstract}
Across almost all scientific disciplines, the instruments that record our experimental data and the methods required for storage and data analysis are rapidly increasing in complexity.
This gives rise to the need for scientific communities to adapt on shorter time scales than traditional university curricula allow for, and therefore requires new modes of knowledge transfer.
The universal applicability of data science tools to a broad range of problems has generated new opportunities to foster exchange of ideas and computational workflows across disciplines.
In recent years, hack weeks have emerged as an effective tool for fostering these exchanges by providing training in modern data analysis workflows.
While there are variations in hack week implementation, all events consist of a common core of three components: tutorials in state-of-the-art methodology, peer-learning and project work in a collaborative environment.
In this paper, we present the concept of a hack week in the larger context of scientific meetings and point out similarities and differences to traditional conferences.
We motivate the need for such an event and present in detail its strengths and challenges.
We find that hack weeks are successful at cultivating collaboration and the exchange of knowledge.
Participants self-report that these events help them both in their day-to-day research as well as their careers.
Based on our results, we conclude that hack weeks present an effective, easy-to-implement, fairly low-cost tool to positively impact data analysis literacy in academic disciplines, foster collaboration and cultivate best practices.
\end{abstract}

\label{sec:introduction}
As data becomes cheaper to gather and store, research across a wide range of disciplines has become increasingly reliant on computational workflows involving a familiarity with aspects of statistical modeling, machine learning, scalable computation, and related skills. In addition, the recent reproducibility crises in several scientific fields has led to the growing realization that improving awareness of open science and reproducibility as well as practical skills in making research reproducible is essential to scientific progress \cite{pashler2012,frye2015,gezelter2015,baker2016}.
Formal university curricula have been relatively slow to offer courses in these important topics: the slack in this area has often been picked-up by extra-curricular, ad-hoc efforts such as workshops (an overview and typography of such efforts in the data science context can be found in \cite{demasi2017}).
Well-known examples are the Software and Data Carpentry workshops providing training in research computing skills through a volunteer instructor program  \cite{b:wilson-swc-lessons-2016,teal2015data}.
At the same time, there has been a rise in the number of domain-specific courses focusing on statistics and computation within their field.
Examples include the \textit{Summer School in Statistics for Astronomers}\footnote{\url{http://astrostatistics.psu.edu/su16/}}, the Google Earth Engine User Summits\footnote{\url{https://events.withgoogle.com/google-earth-engine-user-summit-2017/}}, as well as a variety of project-focused (rather than pedagogical) meetings, such as the dotAstronomy meetings\footnote{\url{http://dotastronomy.com}}.
Shorter, but similar-spirit meetings have been held in conjunction with conferences, such as the Hack Days at the annual American Astronomical Society meetings, the Brainhack hackathons that take place in conjunction with meetings of the Organization for Human Brain Mapping and the Society for Neuroscience\cite{Cameron_Craddock2016-wc}, and a hackathon at the American Geophysical Union meeting\footnote{\url{http://onlinelibrary.wiley.com/doi/10.1002/2014EO480004/pdf}}.
Generally, pedagogically-focused events follow a classic academic model where novices learn new skills from experts, while project-focused workshops emphasize collaborative activities using existing skills.
A disadvantage of the pedagogical model is that it can tends to focus on a one-way flow of information from instructor to student, and can discount the potential contributions by students.
A disadvantage of the project model is the common perception that the week is designed for technical experts, which may discourage others from attending.
In 2014, we initiated an alternative model of ``Hack Weeks'' that try to fill the gaps between these models.
These are week-long events that combine pedagogy (often focused on statistical and computational techniques) together with time for hacks and creative projects, and with the goal of encouraging collaboration and learning among people at various stages of their career.

\begin{figure}
\begin{center}
\includegraphics[width=9cm]{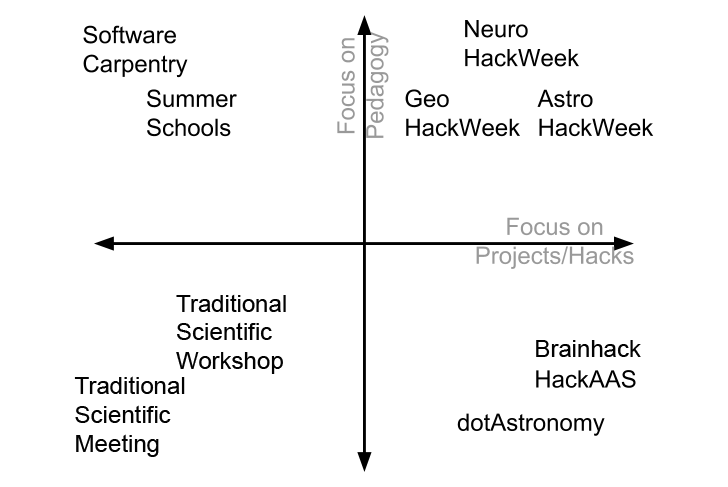}
\caption{Comparison of Extracurricular Workshop Models}
\label{fig:hackspectrum}
\end{center}
\end{figure}

As of the publication of this paper, we have run eight such hack week events: four focused on Astronomy, two focused on Neuroscience, and two focused on Geoscience.
Below we will share some of the philosophy behind the hack week model, practical lessons we have learned in organizing these events, and recommendations for future hack weeks in other disciplines.

\section*{What is a hackathon?}

Hackathons are time-bounded, collaborative events that bring together participants around a shared challenge or learning objective \cite{Decker2015}.
Hackathons originated from the open-source software movement, and have historically focused on software development, particularly in the form of coding sprints, and technology design as a way to motivate innovation, eventually being adopted also within the technology industry.
In recent years, hackathons have expanded into a model for intensive short-term collaboration across disciplinary and topical boundaries.
In addition, because of their focus on participatory engagement, hackathons provide numerous opportunities to `learn by doing' within a constructivist educational framework \cite{Bransford2000-lu,Papert1980-fh}.
With this in mind, hackathons around scientific topics, designed to foster collaboration \cite{Groen2015-cj,Moller2013-ah}, or provide an opportunity to learn \cite{Kienzler2015-zu,Lamers2014-xf}, are becoming more common.

However, in addition to these goals, core element of all hackathons include opportunities for networking, strengthening of social ties, and the building of community connections, both within and across disciplines.
Building on these core elements, there are various implementations of the hackathon concept with respect to the overall purpose, mode of participation, style of work environment and motivation \cite{Drouhard2017}:
``Catalytic'' hackathons seek novel project ideas aimed at solving a tractable, well-defined challenge.
``Contributive'' hackathons seek to improve to an existing effort through focused work on discrete tasks, for example to make up for deficiencies in an ongoing project.
Finally, ``Communal'' hackathons place a strong focus on building a culture of practice and developing resources within an existing community, often defined by a specific domain of knowledge.

Our past hack week events follow most closely the communal hackathon model, as it applies to scientific communities of practice.
Our approach aims to combine structured, tutorial-style instruction with informal education and peer learning opportunities occurring within projects and hacks.
Within the communal model we see these tools being implemented across a spectrum of approaches, the design of which depends on the specific characteristics of each community of practice (Figure \ref{fig:hackspectrum}).
For example, the astronomy community is relatively small and has a foundation of shared approaches and software implementations, allowing for a greater focus on project work over formal tutorials.
In contrast, both the neuro- and geoscience communities covered a broader range of sub-disciplines and had a less cohesive set of existing practices, calling for greater focus on tutorials and education.

We note that the terminology for these events is constantly evolving, and that the ``hackathon'' concept may have implicit connotations that are disfavored in some communities.
One criticism of hackathons is that they propel the ``geek'' stereotype and may present a barrier to creating an inclusive working environment, especially for individuals traditionally underrepresented in science and technology \cite{Decker2015}.
This criticism needs to be actively addressed both through participant selection (see below), and by addressing the possibility that some participants may experience an ``imposter syndrome'' within such circumstances \cite{clance1978imposter}.
Also, while many industry hackathons are competitive, with teams actively competing to solve the same problems for prizes, academic hackathons, and specifically the hack weeks that we have organized are not explicitly competitive.

\section*{Why run a Hack Week?}

There are several reasons to run a hack week of the sort described here.

\begin{itemize}
\item{\textit{Education and Training}: 
While some hack weeks are focused more on education than others (see Figure \ref{fig:hackspectrum}), there is often a skill-development component that entails extensive discussion on reproducible research and open science practices. Participants gain a strong foundation in open science practices from the diverse group setting and go on the become ambassadors for such practices in their respective fields. This type of lateral knowledge transfer is a core attribute of a hack week, and provides an opportunity to learn skills that are not described in papers and software implementations.}

\item{\textit{Tool Development}: Hack weeks present an opportunity for scientific software developers to meaningfully engage with users and critically evaluate applications to particular scientific issues.}

\item{\textit{Community Building}: Hack weeks provide a tremendous opportunity to catalyze community development through a shared interest in solving computational challenges with open source software. These events allow computationally minded researchers to break from the isolation of their academic departments, build connections and spark new collaborations.}

\item{\textit{Interdisciplinary research}: Intensive, time-bounded collaborative events are an excellent opportunity to experiment with concepts, questions, and methods that span boundaries within and across disciplines. Despite the fact that such interdisciplinary experiments are highly impactful \cite{Hall2012-hi}, they are often discouraged in risk averse traditional academia \cite{Sung2003-go, Rhoten2004-fk}}.

\item{\textit{Recruitment and Networking}: Hack weeks are often a melting pot of participants from academia, government, and industry and provide numerous opportunities for networking. Close collaboration in diverse groups exposes skills that might be suitable for careers outside of one's narrow domain.}

\item{\textit{It's fun}: Hack weeks provide a respite from day-to-day research activities and provide a low-stress venue to learn new skills and attempt high-risk projects.}

\end{itemize}

It is worth noting that the reasons for participants to attend a hack week are as diverse as the reasons for running such an event. 
Beginner participants may attend primarily to learn a new technique and may not have an explicit plan for project based work, while others may attend to gain experience in mentoring, or to focusing on an existing project already in progress.

\section*{Audience and Participant Selection}

Hack weeks differ from traditional conferences or summer schools in that knowledge transfer occurs across many levels of seniority, disciplinary boundaries, and novelty of the topics discussed.
In addition, a substantial amount of hack week content is generated during the event itself, requiring active participation from participants.
Therefore in order to maximize learning outcomes and the likelihood for collaborative exchanges, it is crucial that the participant selection process be carried out with considerable care.

In our experience, a participant group that is diverse across categories of diversity, gender, discipline and career track helps to ensure we meet these objectives.
To achieve this diversity, we advocate for a selection process that is as transparent as possible, enabling participants to hold organizers accountable for their selection decisions.
Transparency is necessary for applicants to understand acceptance/rejection decisions, and accountability is of crucial importance for the detection of inherent biases in the selection, which may harm both the event's success as well as the larger community.

One way to maximize transparency in the selection process is to minimize human decision making steps that introduce biases, and to transfer some steps to an algorithm that is easily interpreted, openly available, and can be designed to counter the perpetuation of intrinsic biases. 
We work to achieve this by first assessing the merit of each candidate with respect to the overall goals of the hack week.
We try to minimize bias in this step by blinding ourselves to a candidate's other attributes, including name and other personal information, and assess their candidacy based soley on questions asked specifically for this purpose.
When doing this procedure for a large enough sample, it is unlikely that the resulting pool of acceptable candidates is smaller than the number of available spaces at the workshop.

The second step in the selection procedure then requires tie-breaking between equally acceptable candidates.
It is here where one may impose outside constraints on the selection based on the goals of the workshop.
If multiple competing constraints are considered, this task essentially becomes a complex optimization problem, for which algorithms exist that will outperform any human selection procedure.

One solution to this optimization procedure is implemented in the software \textit{entrofy}\footnote{\url{http://github.com/dhuppenkothen/entrofy}}. The algorithm aims to find a group of participants that together match as closely as possible a requested distribution on specified dimensions (e.g., career stage, geographic location, etc.), to meet pre-set fractions set by the organizers.
For example, organizers may require that half of the participants (or as close as possible to that) be graduate students, while also maximizing the number of different countries from which participants originate.

It is worth noting that this algorithm is vulnerable toward biases in two ways: firstly, humans will set the target fractions for any category of interest.
If these targets reproduce the distribution of the overall sample of candidates, the selection will become essentially random.
Any human biases involved in setting these target fractions will be perpetuated in the selection procedure.
Secondly, perhaps more obviously, the algorithm can only act on information that has been collected.
Biased participant sets may still result from selection procedures that fail to include crucial categories. For example, it would be difficult to produce a student-heavy participant set for a summer school if the algorithm has no information about academic seniority, and impossible to correct gender bias in the pool of applicants, if no information is available about the gender of participants.

\section*{Themes}

To date, all organized hack weeks have been subject-specific, i.e.\ aimed at bringing together a community with a shared scientific interest, such as neuroscience.
Advantages to this approach include shared language and scientific objectives within communities organized by subject, leaving more time for active collaboration on cutting-edge science.
On the other hand, homogeneity may lead to \textit{group think} and inhibit new, creative solutions. 
In this case, it may be advantageous to design a hack week around a technique (e.g.\ Gaussian Processes) or modality (e.g.\ imaging), such as the ImageXD (image processing across domains\footnote{\url{http://http://www.imagexd.org/}} meetings. 
For these events, building a shared vocabulary and understanding of major data analysis problems is crucial, but they also allow for cross-disciplinary diffusion of techniques into other subjects and therefore decrease the risk of duplication of method development efforts.

\section*{Design considerations}

Scheduling, group size and venue are important design considerations contributing to the success of a hack week.   
Longer events allow for a larger taught component, more ambitious projects and cross-disciplinary exchanges. 
By spending more time together, participants are more likely to overcome barriers of professional terminology.
On the other hand, events that are too long may lead to fatigue among attendees, resulting in a drop in positive outcomes later in the workshop.
A well-designed hack week will have a clear schedule that limits the number of parallel sessions, in order to avoid decision fatigue, and will balance the duration of taught components and open project work. 

A hackweek requires a flexible workspace environment with ample opportunity for re-configuration. 
Participants must have access to rooms that can accommodate lectures combined with interactive exchanges and individual work on laptops. 
Workspaces must also accommodate interactive project work where small teams can gather and work together uninterrupted by other groups.    
Universities are an obvious first choice for hosting a hack week given their access to scientific resources, research support and infrastructure. 
However traditional university lecture halls often do not meet the hack week criteria for interactive exchanges.
Fortunately, many universities are experimenting with other types of spaces: for example, the adoption of active learning teaching methods \cite{prince2004} has led to the development of modular classrooms, where seating arrangements can be flexibly modified and group activities can be more readily undertaken. There is a natural tension between keeping the group together and providing physically separated break-out spaces. On the one hand, conducting the workshop in a single large venue improves group cohesion and prevents self-segregation of researchers and ideas. On the other hand, the interactive nature of the workshop may lead to an environment where it is difficult for individuals to focus on the highly complex tasks that are typical for hack week projects. It may hence be advisable to allow for diffusion of the group into adjacent rooms, while providing ample venue for the entire group to congregate for tutorials, breaks and reports.

Another important design consideration is group size.
If the group is too large, chances for random participant exchanges are reduced, and knowledge transfer may decline as the workshop fractures into smaller groups, often among participants who already know each other.
If the group is too small, it is unlikely to achieve the desired level of diversity among participants to foster new collaborations across sub-fields and disciplines.
In the past, we have found groups with sizes between 50 and 70 participants to be large enough to encourage a breadth of projects while allowing the workshop to function as a cohesive group.

As mentioned above, the balance between pedagogy and open project work depends both on the goals of the workshop and the topics around which the workshop is organized.
If participants have little shared knowledge, more teaching may be necessary in order to allow participants to effectively communicate with each other.
In communities where a shared understanding exists, tutorials can be shortened to focus on more advanced or innovative topics, leaving more time for active participation.

Hack week outcomes depend strongly on the interest and engagement of participants.
Some attendees, usually those with a strong background in hackathons and their specific topic of study, arrive with the goal of writing a scientific article.
Other attendees plan to learn a specific topic, such as machine learning, or to analyze a specific data set that relates to the tools covered in the tutorials.
This leads to a wide variety of project types from sandbox-style explorations to focused work efforts.
This breadth of possible outcomes makes it difficult to design for all possible participant goals, and calls for adaptive, flexible leadership among hack week organizers.
The large variety in participant backgrounds and experiences can lead to an increase in the prevalence of impostor syndrome experienced by many participants (see also supplementary materials), and it 
is important to take this into account during workshop design.

\section*{Results}

\begin{figure*}[h!]
\begin{center}
\includegraphics[width=11cm]{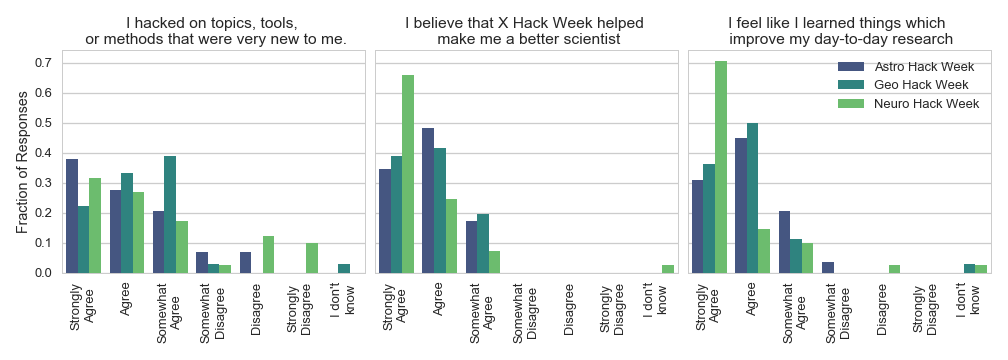}
\includegraphics[width=11cm]{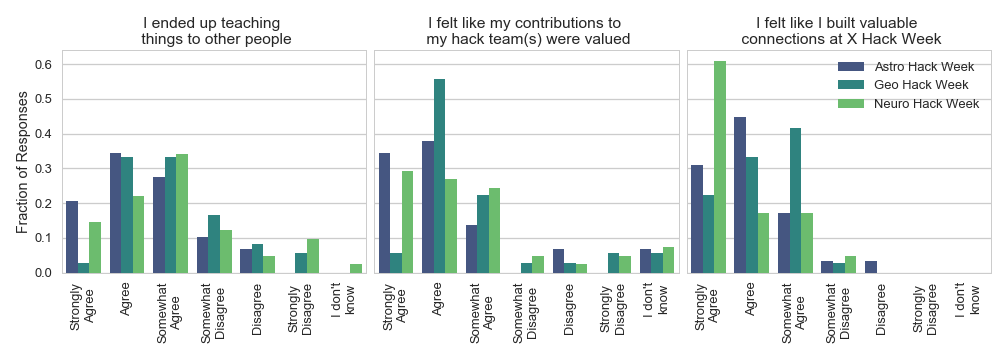}
\includegraphics[width=11cm]{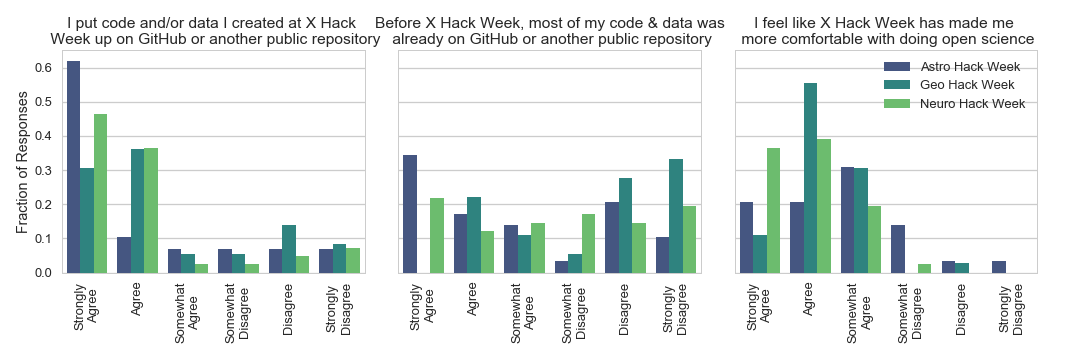}
\caption{{\bf Post-workshop surveys from three hack weeks}: participants in the 2016 astro-, geo- and neuro- hack weeks responded to questions assessing their experiences. We report here about results in three different domains: the development of technical skills (top), collaboration and learning (middle), and shifts in attitudes towards reproducibility and open science (bottom)}
\label{fig:survey}
\end{center}
\end{figure*}

Measuring the success of a hack week objectively is complicated by the variety of objectives that a hack week might have (see above). 
Additionally, the participant-driven, open-ended format facilitates knowledge transfer and collaborations in sometimes surprising ways that escape traditional measures of success.

One key metric is the number of publications that result from hack week projects, but this is a fairly narrow definition of success, in line with standard academic performance indicators.
Assuming that participants work largely in the open during a hack week, and that most projects have a strong programming component another indicator of success is the activity of participants in terms of code written and committed to a public code repository.
Still, these measures ignore learning, community-building as well as networking outcomes, which can be assessed through post-workshop surveys.
Here, we have taken an approach that combines these metrics: we start with survey results, and anecdotally report about publications, new code and projects generated (see the following section).


Focusing on the outcomes of astro-, geo- and neuro- hack weeks (AHW, GHW, NHW, respectively) from 2016, we find that all participants self-report successful learning outcomes in new topics, tools or methods (AHW: 86\%, GHW: 94\%, NHW: 76\% for responses ``somewhat agree'', ``agree'' and ``strongly agree''; Figure \ref{fig:survey}, top panel, left).
The overwhelming majority of respondents at the hack weeks ($>95\%$ for all three events) believed that they learned things that improve their day-to-day research, and that attendance has made them a better scientist ( Figure \ref{fig:survey}, top panel, right and middle).
The majority of participants felt that their contributions to their hack teams was valued, and that they built valuable connections to other researchers (Figure \ref{fig:survey}, middle panel, middle).
This is especially true for Neuro Hack Week, where more than 64\% of participants strongly agreed that they formed valuable connections at NHW (Figure  \ref{fig:survey}, middle panel, right).
Because peer learning is a major mode of knowledge transfer at hack weeks, we asked participants whether they taught other participants.
We find that again a majority agrees with this statement to some degree (AHW: 83\%, GHW: 69\%, NHW: 71\% for combined responses ``somewhat agree'', ``agree'' and ``strongly agree''; Figure \ref{fig:survey}, middle panel, left), though responses are not as unequivocal as they are in some of the other categories.
This is expected: participants new to the field may participate to learn rather than to teach.

We find that the hack weeks have been largely successful at efforts to promote positive attitudes towards reproducibility and open science: at all three events, more than 70\% of all participants (AHW: 79\%, GHW: 72\%, NHW:85\%; Figure \ref{fig:survey}, bottom panel, left) put code or data created at the hack week into a public repository, while a substantially smaller fraction of participants followed this practice before the event (Figure \ref{fig:survey}, bottom panel, middle).
When asking participants whether they had made their code or data openly available in the past, the overall behaviour reflects how general conventions and attitudes differ in different fields.
Astronomy shows the largest degree of openness toward open science, whereas our results indicate that open science is still fairly uncommon in the geosciences, with neuroscience falling in between.
This implies that hack weeks can have the highest impacts in field where a priori engagement in reproducibility efforts is low and significant progress can be made towards changing researchers' attitudes during a collaborative workshop.
Similar attitudes are reflected when asking whether the hack week has made participants more comfortable with open science: again, geoscience shows the large improvement with over 97\% agreeing with this statement to some degree, closely followed by neuroscience (95\%), while there was somewhat less of an impact on participants' attitudes in astronomy (72\%).
Overall, our results show that hack weeks are effective at addressing persisting doubts about making research open and reproducible.

From the very nature of the activities that we encourage in hack weeks, participants in these events produce digital records of their research online. This means that it will be fairly straightforward to evaluate the long-term impact of these activities on participants' productivity (e.g., through contributions to open-source software) in the future. Because all three events are relatively recent, it is still early to evaluate such long-term outcomes, as well as others including publications and collaborations resulting from these events.
There are, however, initial indicators that all hack weeks encouraged long-term engagement with new concepts or tools and that they directly resulted in a number of publications \cite{gullysantiago2015,faria2016,keshavan2017,leonard2017,jordan2017,peterson2017,hahn2017,pricewhelan2017}. For specific examples, see also below.

\subsection{Examples of Hack Week Outcomes}
\label{sec:outcomes}

\subsubsection*{Example 1: Astro Hack Week}
In 2015, a small team used the opportunity of AHW to found a new software project called Stingray\footnote{https://github.com/StingraySoftware/stingray} with the goal of providing well-tested, well-documented implementations of time series analysis algorithms often used in X-ray astronomy.
The start of this project was facilitated by the collaborative environment at Astro Hack Week, including expertise in how to start/run open-source projects, role models of successful projects, and an environment encouraging scientific risk taking. Astro Hack Week enabled participants to seed a new collaboration around a software project needed by the larger community.
Since its beginnings at Astro Hack Week, Stingray has matured into an enduring collaboration within the community with five active maintainers, a number of contributors and four Google Summer of Code projects.
\subsubsection*{Example 2: Geo Hack Week}
In 2016, a GHW project team used Google Earth Engine to explore spatial patterns in climate, topography and population data with the goal of mapping the most suitable locations for renewable energy sites in the United States.
The team used machine learning algorithms in conjunction with the powerful hardware resources provided by Google Earth Engine\footnote{\url{http://georgerichardson.net/2017/04/10/searching-for-energy-in-a-random-forest/}}.
George Richardson, one of the project leads, now works for a renewable resource company in Seattle.
\subsubsection*{Example 3: Neuro Hack Week}
Motion of study participants inside of the MRI machine is a major concern in neuroimaging studies, particularly in studies of children or patients, as they are more likely to move.
During NHW 2016 one of the teams focused on a large and openly available data-set of MRI data from children\footnote{ABIDE: \url{http://preprocessed-connectomes-project.org/abide}}.
To test the effect of motion on the results, the team conducted an analysis in which both the number of experimental subjects included, as well as motion cut-off were varied.
The team (composed of four different researchers from four different institutions in two different countries) continued to work on this project remotely after the end of Neuro Hack Week, and eventually published a paper describing these results in the open access journal Research Ideas and Outcomes \cite{leonard2017}.

\section*{Conclusions}

The fast-paced changes of the computational and methodological landscape require traditional fields of science to rapidly adapt to new data analysis challenges.
Traditional modes of learning, including university curricula, are often too slow to incorporate new developments on short enough time scales to meet their acute need in scientific advancement.
To address this imbalance, new types of workshops, including unconferences, hackathons and bootcamps, have been developed in recent years in various scientific disciplines to exist alongside with and support the existing structure of academic conferences.
Here, we introduce one such concept, hack weeks, and detail the underlying philosophical ideas along with experiences from events held in three different fields

As introduced above, hack weeks serve multiple purposes, including dissemination of state-of-the-art technological advances through the scientific community, building collaborations between academic subdisciplines and fostering interdisciplinary research as well as  promoting open science and reproducibility.
Initial results from three events held in 2016 in three different fields (astronomy, geosciences and neurosciences) indicate that hack weeks succeed at all of these objectives, but that the measure of success is field-specific in that it depends to some degree on how much the concepts hack weeks promote were already adopted within the community.
Hack weeks are still a very young concept, and estimating the long-term impact of these events within the scientific communities they serve will require follow-up over multiple years to asses their effect on collaboration networks, career outcomes for early-career academics and adoption of new methods.
We have shown, however, that hack weeks provide an easy-to-implement, fairly low-cost method to introduce new technologies and methods into scientific fields on much shorter time scales than traditional teaching efforts can.

\section{References}
\bibliographystyle{naturemag}
\bibliography{paper}

\begin{thebibliography}{10}
\expandafter\ifx\csname url\endcsname\relax
  \def\url#1{\texttt{#1}}\fi
\expandafter\ifx\csname urlprefix\endcsname\relax\def\urlprefix{URL }\fi
\providecommand{\bibinfo}[2]{#2}
\providecommand{\eprint}[2][]{\url{#2}}

\bibitem{pashler2012}
\bibinfo{author}{Pashler, H.} \& \bibinfo{author}{Wagenmakers, E.-J.}
\newblock \bibinfo{title}{Editors' introduction to the special section on
  replicability in psychological science}.
\newblock \emph{\bibinfo{journal}{Perspectives on Psychological Science}}
  \textbf{\bibinfo{volume}{7}}, \bibinfo{pages}{528--530}
  (\bibinfo{year}{2012}).
\newblock \urlprefix\url{https://doi.org/10.1177/1745691612465253}.
\newblock \bibinfo{note}{PMID: 26168108},
  \eprint{https://doi.org/10.1177/1745691612465253}.

\bibitem{frye2015}
\bibinfo{author}{Frye, S.~V.} \emph{et~al.}
\newblock \bibinfo{title}{Tackling reproducibility in academic preclinical drug
  discovery}.
\newblock \emph{\bibinfo{journal}{Nature Reviews.Drug Discovery}}
  \textbf{\bibinfo{volume}{14}}, \bibinfo{pages}{733--734}
  (\bibinfo{year}{2015}).
\newblock
  \urlprefix\url{http://ezproxy.library.nyu.edu:2048/login?url=https://search.proquest.com/docview/1766799193?accountid=12768}.
\newblock \bibinfo{note}{Copyright - Copyright Nature Publishing Group Nov
  2015; Last updated - 2016-02-20}.

\bibitem{gezelter2015}
\bibinfo{author}{Gezelter, J.~D.}
\newblock \bibinfo{title}{Open source and open data should be standard
  practices}.
\newblock \emph{\bibinfo{journal}{The Journal of Physical Chemistry Letters}}
  \textbf{\bibinfo{volume}{6}}, \bibinfo{pages}{1168--1169}
  (\bibinfo{year}{2015}).
\newblock \urlprefix\url{http://dx.doi.org/10.1021/acs.jpclett.5b00285}.
\newblock \bibinfo{note}{PMID: 26262967},
  \eprint{http://dx.doi.org/10.1021/acs.jpclett.5b00285}.

\bibitem{baker2016}
\bibinfo{author}{Baker, M.}
\newblock \bibinfo{title}{"1,500 scientists lift the lid on reproducibility"}.
\newblock \emph{\bibinfo{journal}{Nature News Feature}}
  (\bibinfo{year}{2016}).

\bibitem{demasi2017}
\bibinfo{author}{deMasi, O.} \& \bibinfo{author}{Paxton, A.}
\newblock \bibinfo{title}{"complementing the classroom: Advancing data science
  education through ad hoc efforts"}  (\bibinfo{year}{submitted}).

\bibitem{b:wilson-swc-lessons-2016}
\bibinfo{author}{Wilson, G.}
\newblock \bibinfo{title}{Software carpentry: Lessons learned}.
\newblock \emph{\bibinfo{journal}{F1000Research}}  (\bibinfo{year}{2016}).
\newblock \urlprefix\url{http://dx.doi.org/10.12688/f1000research.3-62.v2}.

\bibitem{teal2015data}
\bibinfo{author}{Teal, T.~K.} \emph{et~al.}
\newblock \bibinfo{title}{Data carpentry: workshops to increase data literacy
  for researchers}.
\newblock \emph{\bibinfo{journal}{International Journal of Digital Curation}}
  \textbf{\bibinfo{volume}{10}}, \bibinfo{pages}{135--143}
  (\bibinfo{year}{2015}).

\bibitem{Cameron_Craddock2016-wc}
\bibinfo{author}{Cameron~Craddock, R.} \emph{et~al.}
\newblock \bibinfo{title}{Brainhack: a collaborative workshop for the open
  neuroscience community}.
\newblock \emph{\bibinfo{journal}{Gigascience}} \textbf{\bibinfo{volume}{5}},
  \bibinfo{pages}{16} (\bibinfo{year}{2016}).

\bibitem{Decker2015}
\bibinfo{author}{Decker, A.}, \bibinfo{author}{Eiselt, K.} \&
  \bibinfo{author}{Voll, K.}
\newblock \bibinfo{title}{Understanding and improving the culture of
  hackathons: Think global hack local}.
\newblock \emph{\bibinfo{journal}{2015 IEEE Frontiers in Education Conference
  (FIE)}} \textbf{\bibinfo{volume}{00}}, \bibinfo{pages}{1--8}
  (\bibinfo{year}{2015}).

\bibitem{Bransford2000-lu}
\bibinfo{author}{Bransford, J.~D.}, \bibinfo{author}{Brown, A.~L.},
  \bibinfo{author}{Cocking, R.~R.} \& \bibinfo{author}{{Others}}.
\newblock \bibinfo{title}{How people learn} (\bibinfo{year}{2000}).

\bibitem{Papert1980-fh}
\bibinfo{author}{Papert, S.}
\newblock \bibinfo{title}{Mindstorms. {NY}} (\bibinfo{year}{1980}).

\bibitem{Groen2015-cj}
\bibinfo{author}{Groen, D.} \& \bibinfo{author}{Calderhead, B.}
\newblock \bibinfo{title}{Science hackathons for developing interdisciplinary
  research and collaborations}.
\newblock \emph{\bibinfo{journal}{Elife}} \textbf{\bibinfo{volume}{4}},
  \bibinfo{pages}{e09944} (\bibinfo{year}{2015}).

\bibitem{Moller2013-ah}
\bibinfo{author}{M{\"{o}}ller, S.} \emph{et~al.}
\newblock \bibinfo{title}{Sprints, hackathons and codefests as community gluons
  in computational biology}.
\newblock \emph{\bibinfo{journal}{EMBnet.journal}}
  \textbf{\bibinfo{volume}{19}}, \bibinfo{pages}{40--42}
  (\bibinfo{year}{2013}).

\bibitem{Kienzler2015-zu}
\bibinfo{author}{Kienzler, H.}
\newblock \bibinfo{title}{Bringing students into research by hacking global
  health}.
\newblock \emph{\bibinfo{journal}{Higher Education Research Network Journal}}
  \bibinfo{pages}{17} (\bibinfo{year}{2015}).

\bibitem{Lamers2014-xf}
\bibinfo{author}{Lamers, M.~H.}, \bibinfo{author}{van~der Putten, P.} \&
  \bibinfo{author}{Verbeek, F.~J.}
\newblock \bibinfo{title}{Observations on tinkering in scientific education}.
\newblock In \emph{\bibinfo{booktitle}{Entertaining the Whole World}},
  Human--Computer Interaction Series, \bibinfo{pages}{137--145}
  (\bibinfo{publisher}{Springer London}, \bibinfo{year}{2014}).

\bibitem{Drouhard2017}
\bibinfo{author}{Drouhard, M.}, \bibinfo{author}{Tanweer, A.} \&
  \bibinfo{author}{Fiore-Gartland, B.}
\newblock \bibinfo{title}{"a typology of hackathon events"}
  (\bibinfo{year}{2017}).

\bibitem{clance1978imposter}
\bibinfo{author}{Clance, P.~R.} \& \bibinfo{author}{Imes, S.~A.}
\newblock \bibinfo{title}{The imposter phenomenon in high achieving women:
  Dynamics and therapeutic intervention.}
\newblock \emph{\bibinfo{journal}{Psychotherapy: Theory, Research \& Practice}}
  \textbf{\bibinfo{volume}{15}}, \bibinfo{pages}{241} (\bibinfo{year}{1978}).

\bibitem{Hall2012-hi}
\bibinfo{author}{Hall, K.~L.} \emph{et~al.}
\newblock \bibinfo{title}{Assessing the value of team science: a study
  comparing center- and investigator-initiated grants}.
\newblock \emph{\bibinfo{journal}{Am. J. Prev. Med.}}
  \textbf{\bibinfo{volume}{42}}, \bibinfo{pages}{157--163}
  (\bibinfo{year}{2012}).

\bibitem{Sung2003-go}
\bibinfo{author}{Sung, N.~S.} \emph{et~al.}
\newblock \bibinfo{title}{Science education. educating future scientists}.
\newblock \emph{\bibinfo{journal}{Science}} \textbf{\bibinfo{volume}{301}},
  \bibinfo{pages}{1485} (\bibinfo{year}{2003}).

\bibitem{Rhoten2004-fk}
\bibinfo{author}{Rhoten, D.} \& \bibinfo{author}{Parker, A.}
\newblock \bibinfo{title}{Education. risks and rewards of an interdisciplinary
  research path}.
\newblock \emph{\bibinfo{journal}{Science}} \textbf{\bibinfo{volume}{306}},
  \bibinfo{pages}{2046} (\bibinfo{year}{2004}).

\bibitem{prince2004}
\bibinfo{author}{Prince, M.}
\newblock \bibinfo{title}{Does active learning work? a review of the research}.
\newblock \emph{\bibinfo{journal}{Journal of Engineering Education}}
  \textbf{\bibinfo{volume}{93}}, \bibinfo{pages}{223--231}
  (\bibinfo{year}{2004}).
\newblock \urlprefix\url{http://dx.doi.org/10.1002/j.2168-9830.2004.tb00809.x}.

\bibitem{gullysantiago2015}
\bibinfo{author}{{Gully-Santiago}, M.}, \bibinfo{author}{{Jaffe}, D.~T.} \&
  \bibinfo{author}{{White}, V.}
\newblock \bibinfo{title}{{Optical characterization of gaps in directly bonded
  Si compound optics using infrared spectroscopy}}.
\newblock \emph{\bibinfo{journal}{ao}} \textbf{\bibinfo{volume}{54}},
  \bibinfo{pages}{10177} (\bibinfo{year}{2015}).
\newblock \eprint{1511.01183}.

\bibitem{faria2016}
\bibinfo{author}{{Faria}, J.~P.} \emph{et~al.}
\newblock \bibinfo{title}{{Uncovering the planets and stellar activity of
  CoRoT-7 using only radial velocities}}.
\newblock \emph{\bibinfo{journal}{aap}} \textbf{\bibinfo{volume}{588}},
  \bibinfo{pages}{A31} (\bibinfo{year}{2016}).
\newblock \eprint{1601.07495}.

\bibitem{keshavan2017}
\bibinfo{author}{Keshavan, A.} \emph{et~al.}
\newblock \bibinfo{title}{Mindcontrol: A web application for brain segmentation
  quality control}.
\newblock \emph{\bibinfo{journal}{NeuroImage}}  (\bibinfo{year}{2017}).
\newblock
  \urlprefix\url{http://www.sciencedirect.com/science/article/pii/S1053811917302707}.

\bibitem{leonard2017}
\bibinfo{author}{Leonard, J.}, \bibinfo{author}{Flournoy, J.},
  \bibinfo{author}{de~los Angeles, C. P.~L.} \& \bibinfo{author}{Whitaker, K.}
\newblock \bibinfo{title}{How much motion is too much motion? determining
  motion thresholds by sample size for reproducibility in developmental
  resting-state mri}.
\newblock \emph{\bibinfo{journal}{Research Ideas and Outcomes}}
  \textbf{\bibinfo{volume}{3}}, \bibinfo{pages}{e12569} (\bibinfo{year}{2017}).
\newblock \urlprefix\url{https://doi.org/10.3897/rio.3.e12569}.
\newblock \eprint{https://doi.org/10.3897/rio.3.e12569}.

\bibitem{jordan2017}
\bibinfo{author}{Jordan, K.~M.}, \bibinfo{author}{Keshavan, A.},
  \bibinfo{author}{Mandelli, M.~L.} \& \bibinfo{author}{Henry, R.~G.}
\newblock \bibinfo{title}{Cluster-viz: A tractography qc tool}.
\newblock \emph{\bibinfo{journal}{Research Ideas and Outcomes}}
  \textbf{\bibinfo{volume}{3}}, \bibinfo{pages}{e12394} (\bibinfo{year}{2017}).
\newblock \urlprefix\url{https://doi.org/10.3897/rio.3.e12394}.
\newblock \eprint{https://doi.org/10.3897/rio.3.e12394}.

\bibitem{peterson2017}
\bibinfo{author}{Peterson, D.}
\newblock \bibinfo{title}{Streamlining the process of 3d printing a brain from
  a structural mri}.
\newblock \emph{\bibinfo{journal}{Research Ideas and Outcomes}}
  \textbf{\bibinfo{volume}{3}}, \bibinfo{pages}{e13394} (\bibinfo{year}{2017}).
\newblock \urlprefix\url{https://doi.org/10.3897/rio.3.e13394}.
\newblock \eprint{https://doi.org/10.3897/rio.3.e13394}.

\bibitem{hahn2017}
\bibinfo{author}{{Hahn}, C.} \emph{et~al.}
\newblock \bibinfo{title}{{Approximate Bayesian computation in large-scale
  structure: constraining the galaxy-halo connection}}.
\newblock \emph{\bibinfo{journal}{mnras}} \textbf{\bibinfo{volume}{469}},
  \bibinfo{pages}{2791--2805} (\bibinfo{year}{2017}).
\newblock \eprint{1607.01782}.

\bibitem{pricewhelan2017}
\bibinfo{author}{{Price-Whelan}, A.~M.}, \bibinfo{author}{{Hogg}, D.~W.},
  \bibinfo{author}{{Foreman-Mackey}, D.} \& \bibinfo{author}{{Rix}, H.-W.}
\newblock \bibinfo{title}{{The Joker: A Custom Monte Carlo Sampler for
  Binary-star and Exoplanet Radial Velocity Data}}.
\newblock \emph{\bibinfo{journal}{apj}} \textbf{\bibinfo{volume}{837}},
  \bibinfo{pages}{20} (\bibinfo{year}{2017}).
\newblock \eprint{1610.07602}.

\end{thebibliography}


\begin{addendum}
 \item{The authors would like to thank Laura Nor\'{e}n (NYU) for help on ethics and IRB, Stuart Geiger for helping to formulate the questionnaires that served as the basis for the results presented here, Brittany Fiore-Gartland and Jason Yeatman for comments on the manuscript, and Tal Yarkoni for advice regarding automated selection procedures. This work was partially supported by the Moore-Sloan Data Science Environments at UC Berkeley, New York University, and the University of Washington. Neuro hack week is supported through a grant from the National Institute for Mental Health (1R25MH112480). Daniela Huppenkothen is partially supported by the James Arthur Postdoctoral Fellowship at NYU.}
 \item[Competing Interests]{The authors declare that they have no competing financial interests.}
 \item[Correspondence]{Correspondence and requests for materials should be addressed to D.Huppenkothen.~(email: dhuppenk@uw.edu).}
\end{addendum}



\end{document}